\documentclass[conference]{IEEEtran}

\usepackage{epsfig}
\usepackage{graphicx}
\usepackage{amsmath}
\usepackage{amssymb}
\usepackage{amsxtra}
\usepackage{here}
\usepackage{rawfonts}
\usepackage{times}
\usepackage{url}
\usepackage{cite}
\usepackage{ dsfont }

\usepackage{cases}
\usepackage{adjustbox}

\usepackage{slashbox}
\usepackage{epstopdf}

\usepackage{cite}
\newtheorem{prop}{\bf Proposition}

\newtheorem{definition}{\bf Definition}

\newlength{\aligntop}
\newlength{\alignbot}
\makeatletter

\renewenvironment{align}{%
  \vspace{\aligntop}
  \start@align\@ne\st@rredfalse\m@ne
}{%
  \math@cr \black@\totwidth@
  \egroup
  \ifingather@
    \restorealignstate@
    \egroup
    \nonumber
    \ifnum0=`{\fi\iffalse}\fi
  \else
    $$%
  \fi
  \ignorespacesafterend%
  \vspace{\alignbot}\par\noindent
}

\IEEEoverridecommandlockouts

\begin{document}
\title{\huge Prospect Theory for Enhanced Cyber-Physical Security of Drone Delivery Systems: A Network Interdiction Game\vspace{-0.55cm}}
\author{\IEEEauthorblockN{Anibal Sanjab$^1$, Walid Saad$^1$, and Tamer Ba\c{s}ar$^2$} \IEEEauthorblockA{\small
$^1$ Wireless@VT, Bradley Department of Electrical and Computer Engineering, Virginia Tech, Blacksburg, VA, USA,\\
 Emails: \url{{anibals,walids}@vt.edu}\\
 $^2$ Coordinated Science Laboratory, University of Illinois at Urbana-Champaign, IL, USA, Email: \url{basar1@illinois.edu}\vspace{-0.6cm}
 }%
\thanks{This research was supported by the U.S. National Science Foundation under Grants CNS-1446621 and ACI-1541105.}
    }
\date{}
\maketitle

\begin{abstract}
The use of unmanned aerial vehicles (UAVs) as delivery systems of online goods is rapidly becoming a global norm, as corroborated by Amazon's ``Prime Air'' and  Google's ``Project Wing'' projects. However, the real-world deployment of such drone delivery systems faces many cyber-physical security challenges. In this paper, a novel mathematical framework for analyzing and enhancing the security of drone delivery systems is introduced. In this regard, a zero-sum network interdiction game is formulated between a vendor, operating a drone delivery system, and a malicious attacker. In this game, the vendor seeks to find the optimal path that its UAV should follow, to deliver a purchase from the vendor's warehouse to a customer location, to minimize the delivery time. Meanwhile, an attacker seeks to choose an optimal location to interdict the potential paths of the UAVs, so as to inflict cyber or physical damage to it, thus, maximizing its delivery time. First, the Nash equilibrium point of this game is characterized. Then, to capture the subjective behavior of both the vendor and attacker, new notions from prospect theory are incorporated into the game. These notions allow capturing the vendor's and attacker's i) subjective perception of attack success probabilities, and ii) their disparate subjective valuations of the achieved delivery times relative to a certain target delivery time. 
Simulation results have shown that the subjective decision making of the vendor and attacker leads to adopting risky path selection strategies which inflict delays to the delivery, thus, yielding unexpected delivery times which surpass the target delivery time set by the vendor.                  
\end{abstract}
\vspace{-0.25cm}
\section{Introduction}
Drone and unmanned aerial vehicle (UAV)-assisted delivery systems are rapidly moving from fiction to reality~\cite{UAVVTCnn,Amazon}. Recent examples include Google's ``Project Wing'' that has recently tested its drones as part of a food delivery system at Virginia Tech~\cite{UAVVTCnn} and the ``Amazon Prime Air'' program through which online shoppers will be given the opportunity of choosing UAV delivery as a $30$-minutes delivery option~\cite{Amazon}.  

However, the implementation of such drone delivery systems is faced with a set of technical challenges ranging from optimized navigation~\cite{DroneDeliveryLifeguardRing,DroneDeliveryDrugShipment,DroneDeliverySystemCollisionAvoidance} to UAV control and system security. In particular, drone delivery systems are vulnerable to a myriad of cyber and physical attacks. On the physical side, to avoid conflict with manned aviations, the altitude of UAVs is limited to around 400 ft~\cite{Amazon} which puts them in the range of civilian-owned hunting rifles which can target them~\cite{SniperSGAttack}. 
Moreover, on the cyber side, UAVs are vulnerable to a range of cyber threats targeting their communication links with ground control, as well as with other air units~\cite{UAVCyberSec1,UAVCyberSec2,UAVCyberSec3}. 
In fact, a number of recent works have characterized a handful of cybersecurity threats against UAVs~\cite{UAVCyberSec1,UAVCyberSec2,UAVCyberSec3}. For example, the work in~\cite{UAVCyberSec1} provided a general overview of cyber attacks which can target the confidentiality, integrity, and availability of UAV systems. The work in~\cite{UAVCyberSec2} focused on the security of the communications links between ground control and unmanned aircrafts, exposing the underlying threats on UAV controllability and operation. Moreover, the work in~\cite{UAVCyberSec3} provided a demonstration in which the authors successfully launched data injection attacks against a typical UAV used by law enforcement agencies for critical applications. 

Even though UAVs used in drone delivery systems are also subject to such physical~\cite{SniperSGAttack} and cyber attacks~\cite{UAVCyberSec1,UAVCyberSec2,UAVCyberSec3}, the emerging research on drone delivery systems\cite{DroneDeliveryLifeguardRing,DroneDeliveryDrugShipment,DroneDeliverySystemCollisionAvoidance}, surprisingly, merely focuses on enhancing the efficiency and precision of the UAVs, and enabling their integration with manned aviation, with little to no focus on addressing and analyzing the underlying security challenges. Moreover, the works in \cite{UAVCyberSec1,UAVCyberSec2,UAVCyberSec3} are either qualitative or focused on isolated military drone experiments that do not necessarily capture the cyber-physical security threats in drone delivery systems. \emph{To the best of our knowledge, no work has focused on studying the security of UAV delivery systems against cyber-physical attacks}.  

The main contribution of this paper is to develop the first comprehensive analysis of the cyber-physical security of drone-based delivery systems. In particular, we consider a zero-sum network interdiction game between a vendor (delivering purchases via UAVs) and a malicious attacker. In this game, the vendor, referred to as an evader, seeks to choose the optimal path strategy for its UAV, from the warehouse to a customer location, to evade attacks along the way and minimize its expected delivery time. On the other hand, the attacker or interdictor, aims at choosing the optimal attack locations along the paths traversed by the UAV to interdict the UAV, causing cyber or physical damage, with the goal of maximizing the delivery time. We then show that this network interdiction game is equivalent to a zero-sum matrix game whose Nash equilibrium (NE) can be derived by solving two linear programming (LP) problems. We then prove that the value functions of these LPs can be directly used to compute the expected delivery time under the NE strategies. 

Moreover, to capture the potential subjective behavior of the vendor and the attacker, we incorporate tools from \emph{prospect theory} (PT)~\cite{Kahneman1979Prospect}, a Nobel prize-winning decision theory, in the game formulation. The incorporation of PT enables modeling i) the subjective perceptions of the likelihood of a potential attack to be successful, and ii) the subjective assessment of an achieved delivery time, pertaining to the vendor and attacker. In fact, the merit of a drone delivery system lies in its ability of meeting a certain short target delivery time (e.g. 30 mins for Amazon Prime Air~\cite{Amazon}, or shorter and stricter time restrictions for medical and safety applications~\cite{DroneDeliveryLifeguardRing,DroneDeliveryDrugShipment}). Thus, an achieved delivery time is naturally assessed with respect to this target delivery time rather than as an absolute quantity. As such, using PT, we model the way in which each of the players, vendor and attacker, subjectively values delivery time relative to their reference point which can represent, for example, the target delivery time. 
Our results show that such subjective perceptions and decision making, of the vendor and attacker, lead to the adoption of risky path selection strategies which cause delays to the delivery and, hence, lead to delivery times which can exceed the target delivery time set by the vendor. 

The rest of the paper is organized as follows. Section~\ref{sec:NetworkModel} presents our system model. Section~\ref{sec:Game} introduces the formulated network interdiction game while Section~\ref{sec:Prospect} discusses the PT version of this game. Section~\ref{sec:NumRes} presents a number of numerical results while Section~\ref{sec:Conclusion} concludes the paper.              

\section{System Model}\label{sec:NetworkModel}
\begin{figure}[t!]
  \begin{center}
   \vspace{-0.35cm}
    \includegraphics[width=5.5cm]{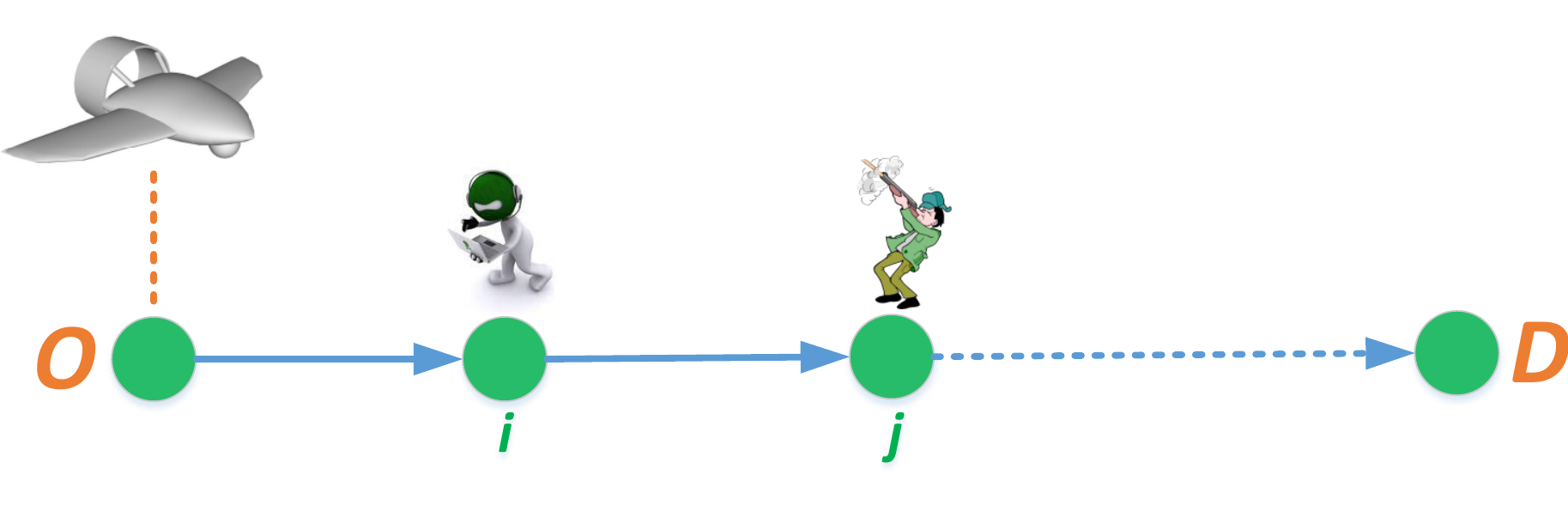}
    \vspace{-0.4cm}
    \caption{\label{fig:OtoD} Threat points from warehouse $(O)$ to customer location $(D)$.}
  \end{center}\vspace{-0.9cm}
\end{figure}

Consider a drone delivery system in which a UAV is used for delivering online purchases, similarly to the paradigms of by Amazon~\cite{Amazon} and Project Wing~\cite{UAVVTCnn}.  
Once an online order is placed, the vendor will schedule its UAV to deliver the product from a warehouse location, $O$, to the customer's delivery location, $D$. The goal of the vendor is to minimize the delivery time (and transportation cost) and, hence, it chooses the shortest path from $O$ to $D$. However, as shown in Fig.~\ref{fig:OtoD}, an adversary might be located at a number of locations or ``danger points'' (such as $i$ and $j$ in Fig.~\ref{fig:OtoD}), along this path aiming to launch a cyber or physical attack targeting the UAV. 
A successful attack leads to the destruction of the UAV which then requires re-sending a similar product from $O$ to $D$, therefore, incurring significant delays. Thus, the expected delivery time of the package is directly dependent on the probability of a successful attack on the UAV along its delivery path. To minimize the expected delivery time, rather than solely considering the shortest physical path, the vendor can consider alternative paths which can potentially decrease the expected delivery time. Such alternative paths may also include danger points.  

\begin{figure}[t!]
  \begin{center}
   \vspace{-0.35cm}
    \includegraphics[width=6.5cm]{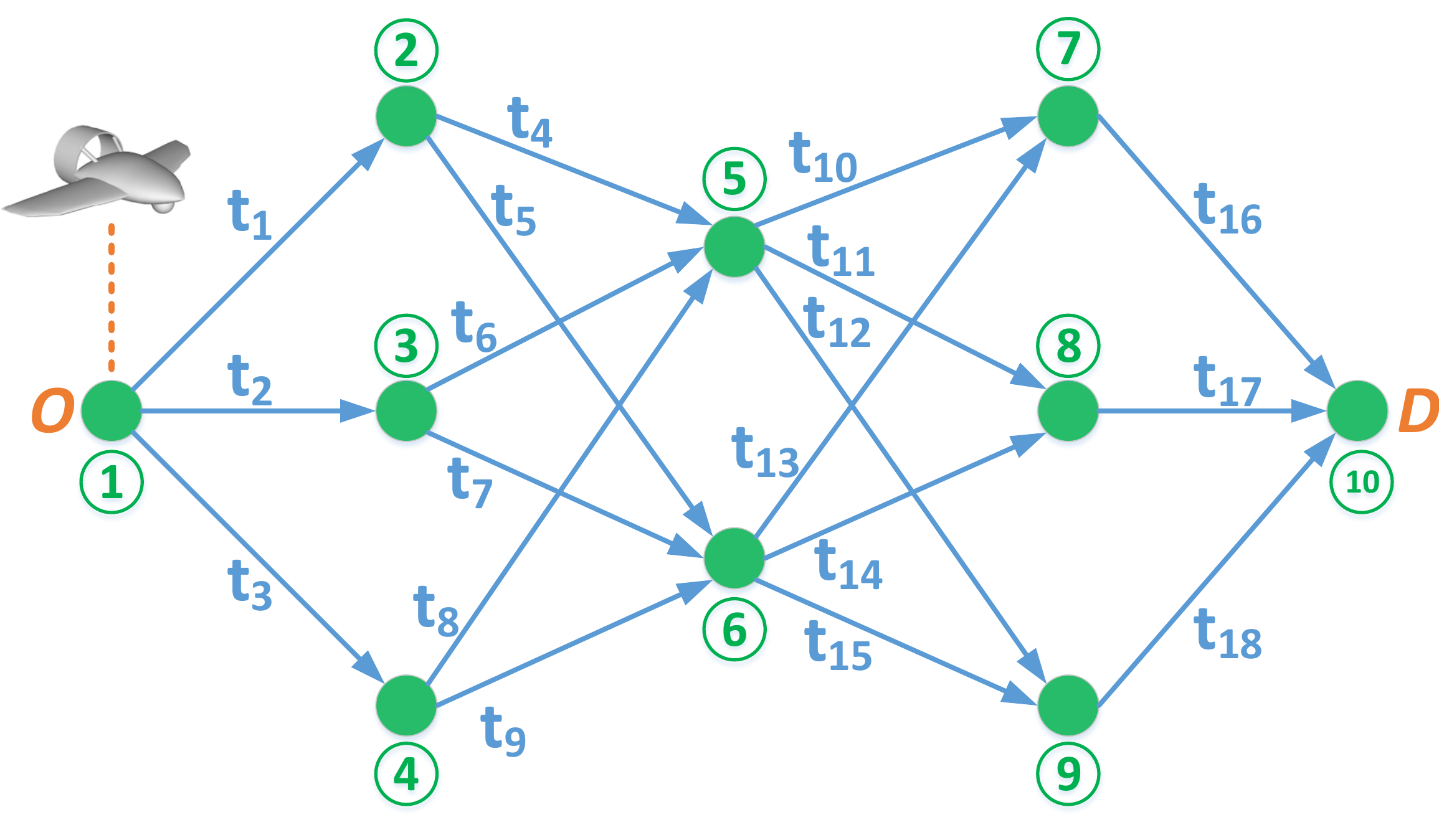}
    \vspace{-0.4cm}
    \caption{\label{fig:Network} Warehouse-to-customer security graph.}
  \end{center}\vspace{-0.9cm}
\end{figure}

The set of danger points represents inevitable locations, situated along the possible paths from $O$ to $D$, from which attacks can be carried out. Such locations can represent points that expose the UAV, such as high hills or high buildings, located between $O$ and $D$. These high locations represent a source of threat since they allow a line-of-sight between the attacker and the UAV as well as spatial proximity. Thus, they enable targeting a traversing UAV with physical (such as shooting the UAV) and cyber (such as jamming) attacks.
To model the possible delivery paths from $O$ to $D$, we consider an $O$ to $D$ network represented by a directed graph $\mathcal{G}(\mathcal{N},\mathcal{E})$ as shown in Fig.~\ref{fig:Network}. In this graph, $\mathcal{N}$ is the set of $N$ nodes, or vertices which represent the danger points between $O$ and $D$, and $\mathcal{E}$ is the set of $E$ edges.  

In practice, given that the UAV may not be limited by predefined airways\footnote{Our system can still accommodate a future case in which the UAV may be regulated to a small set of paths.}, there can be an infinite number of paths connecting $O$ and $D$. Each such path will include a subset of danger points, and different paths might share common danger points. Thus, from a security perspective, this large set of possible paths can be captured by the set of danger points that each path traverses. 
Now, considering each two neighboring danger points, such as nodes 3 and 5 in Fig.~\ref{fig:Network}, there can exist an infinite number of ways in which a drone can move from point $m$ to point $n$. However, given that the vendor aims at minimizing the delivery time, the infinite set of edges connecting $m$ to $n$ can only be represented by the shortest edge between the two vertices. Thus, the graph $\mathcal{G}$ includes only the shortest paths between each two danger points. 
As such, $\mathcal{G}(\mathcal{N},\mathcal{E})$ is a security model which represents the continuous geographical space between $O$ and $D$ in terms of its danger points and the shortest edges connecting them. Moreover, if edge $e_k\in\mathcal{E}$ connects two neighboring danger points $m$ and $n$, we let $t_k$ be the time needed by the UAV to travel from $m$ to $n$ over $e_k$. We also let     
$p_n$ be the probability with which an attack launched from a location $n\in\mathcal{N}$ is successful.

Let $\mathcal{H}$ be the set of $H$ simple paths (with no repeated vertices) from $O$ to $D$ over the graph $\mathcal{G}$. In this respect, $h\in\mathcal{H}$ is a sequence of nodes and edges which connect $O$ to $D$. Within each $h$, the nodes and edges are unique, and thus, we represent $h$ by its sequence of traversed nodes. Each path $h\in\mathcal{H}$, hence, constitutes a subset of $\mathcal{N}$. For example, $h_1\triangleq (1,2,5,7,10)$ is a path from $O$ to $D$ in Fig.~\ref{fig:Network}. To this end, we define the path-node incidence matrix $\boldsymbol{L}$ of size $(H\times N)$ with each element $l_{hn}$, $\forall h\in\mathcal{H}\,,\forall n\in\mathcal{N}$, being: $l_{hn}=1$ if $n\in h$, and $l_{hn}=0$ otherwise. 
In addition, we define, $f^h(.)\!: h\rightarrow\mathds{R}$, as a distance function over path $h\in\mathcal{H}$ which takes as input a node $n\in h$ and returns the time to reach this node $n$ from the origin $O$, following path $h$. For example, in Fig.~\ref{fig:Network}, $f^{h_1}(5)=t_1+t_4$ where $h_1\triangleq (1,2,5,7,10)$.

Given graph $\mathcal{G}$, the vendor, $U$, is an \emph{evader} aiming at choosing an optimal path for its UAV from $O$ to $D$ to evade the attack and minimize the expected delivery time, $T$. On the other hand, the adversary, $A$, is an \emph{interdictor} which aims at choosing a node (a danger point) from which to launch the attack and interdict the path of the UAV while maximizing $T$. As such, we model the decision making processes of $U$ and $A$ using a zero-sum \emph{network interdiction game}~\cite{NetworkInterdictionGame} on graph $\mathcal{G}$ as defined next. 
\section{Game Formulation and Solution Method}\label{sec:Game}

\subsection{Game Formulation}\label{subsec:GameFormulation}
In the proposed game, the vendor aims at choosing an optimal probability distribution $\boldsymbol{y}\triangleq [y_1,y_2,...,y_H]^T\in\mathcal{Y}$ over the set of possible paths, $\mathcal{H}$, from $O$ to $D$ (mixed path-selection strategy) where $\mathcal{Y}=\{\boldsymbol{y}\in\mathds{R}^H: \boldsymbol{y}\geq0, \sum_{h=1}^{H} y_h=1\}.$ 
Randomizing between its path selection strategies is beneficial to the vendor so as not to make it trivial for the adversary to guess the admitted path. A similar randomization logic can be used by the attacker. Indeed, the attacker will choose an optimal probability distribution $\boldsymbol{x}\triangleq [x_1,x_2,...,x_N]^T \in\mathcal{X}$ over the set, $\mathcal{N}$, of possible attack locations (i.e. mixed interdiction-location strategy) where
$\mathcal{X}=\{\boldsymbol{x}\in\mathds{R}^N: \boldsymbol{x}\geq0, \sum_{n=1}^{N} x_n=1\}.$ 

For an attacker located at node $n$ (along the path of the UAV), the attack's success probability is given by $p_n$. When the attack is successful, the UAV is interdicted/destroyed and, hence, a new item needs to be sent from $O$. Thus, when the UAV reaches a certain $n\in h$, it may continue its path $h$ unaffected with probability $(1-p_n)$ or it will be destroyed with probability $p_n$ which is equivalent to being sent back to the starting point $O$. Here, we assume that, when the item is re-sent, the path $h$ which was taken at the first attempt will be cleared from potential threats by law enforcement agencies (due to the reported previous security breach) thus allowing $U$ to safely send a replacement along path $h$ without the threat of any interdiction\footnote{The relaxation of this assumption can be performed through an alternative \emph{repeated game} formulation in which at each time the UAV is successfully attacked, the whole game repeats. This extension will be subject of future work.}. Given that $y_h$ is the probability with which the vendor chooses path $h\in\mathcal{H}$ and $x_n$ is the probability with which the attacker chooses location $n\in\mathcal{N}$, the expected delivery time, $T$, will be:
\begin{flalign}\label{eq:ExpectedDeliveryTime}
T&=\sum_{h\in\mathcal{H}}\sum_{n\in\mathcal{N}}y_hx_n[l_{hn}p_n(f^h(n)\textrm{$+$}f^h(D))\textrm{$+$}(1\textrm{$-$}l_{hn}p_n)f^h(D)]&\nonumber\\
&=\sum_{h\in\mathcal{H}}\sum_{n\in\mathcal{N}}y_hx_n[l_{hn}p_n f^h(n)+ f^h(D)].&
\end{flalign}

In this respect, we define an $(H\times N)$ matrix $\boldsymbol{M}$ whose elements are $m_{hn}=l_{hn}p_n f^h(n)+ f^h(D) \>\forall h\in\mathcal{H}\textrm{ and }n\in\mathcal{N}.$  

Therefore, the expected delivery time can be defined as:
\begin{align}
T=\boldsymbol{y}^T\boldsymbol{M}\boldsymbol{x}.
\end{align}

Given that the objective of the vendor is to minimize $T$ while that of the attacker is to maximize $T$, the vendor's problem can be formulated as a min-max problem $(P_1)$ as follows:
\begin{flalign}
\textrm{$(P_1)$:}\>\>\>\>\>\>&T^*=\min_{\boldsymbol{y}}\max_{\boldsymbol{x}} \boldsymbol{y}^T\boldsymbol{M}\boldsymbol{x},& \label{eq:UAVProblem}\\
\textrm{s.t.}\>\>\>&\boldsymbol{1}_N\boldsymbol{x}=1,\, \boldsymbol{1}_H \boldsymbol{y}=1,&\label{eq:Sumx}\\
&\boldsymbol{x}\geq 0,\,\boldsymbol{y}\geq0,&\label{eq:Posy}
\end{flalign}
where $\boldsymbol{1}_N\triangleq [1,...,1]^T\in\mathds{R}^N$ and $\boldsymbol{1}_H\triangleq [1,...,1]^T\in\mathds{R}^H$. The set of constraints of $(P_1)$ are equivalent to restricting $\boldsymbol{x}$ and $\boldsymbol{y}$ to $\boldsymbol{x}\in\mathcal{X}$ and $\boldsymbol{y}\in\mathcal{Y}$.  
The attacker's problem can be expressed as the max-min counterpart of $(P_1)$. 

The choice of $\boldsymbol{y}$ and $\boldsymbol{x}$ following, respectively, the min-max problem $(P_1)$ and the max-min problem introduced later in~(\ref{eq:AProblem}) constitutes the selection of what is known as \emph{security strategies}~\cite{GT01}. Security strategies are common when studying security problems~\cite{SanjabSaadJournal} since they consider the opponent to inflict worst-case scenarios. For example, in the min-max formulation in~(\ref{eq:UAVProblem}), the vendor considers that the attacker's response to any path strategy $\boldsymbol{y}$ consists of choosing the attack strategy $\boldsymbol{x}\in\mathcal{X}$ which will lead to the highest possible expected delivery time (worst-case scenario to the vendor). 

\subsection{Solution Method}\label{subsec:SolutionMethod}
By inspecting ~(\ref{eq:UAVProblem}), one can see that the maximization is carried out as a function of a given $\boldsymbol{y}$, i.e., the choice of optimal $\boldsymbol{x}\in\mathcal{X}$ can depend on $\boldsymbol{y}$. As such,~(\ref{eq:UAVProblem}) can be written as: 
$\min_{\boldsymbol{y}\in\mathcal{Y}}u_1(\boldsymbol{y}),$ 
where 
$u_1(\boldsymbol{y})=\max_{\boldsymbol{x}\in\mathcal{X}} \boldsymbol{y}^T\boldsymbol{M}\boldsymbol{x}\geq \boldsymbol{y}^T\boldsymbol{M}\boldsymbol{x} \,\, \forall \boldsymbol{x}\in\mathcal{X}.$ 
By definition of $\mathcal{X}$ as an $N$-dimensional simplex, the last inequality can be stated as follows:
\begin{align}\label{eq:MaxAloneTranformed}
\boldsymbol{M}^T\boldsymbol{y}\leq\boldsymbol{1}_N u_1(\boldsymbol{y}).
\end{align}

By performing the change of variables $\hat{\boldsymbol{y}}=\boldsymbol{y}/u_1(\boldsymbol{y})$, the min-max problem, $(P_1)$, can be re-formulated as a linear programming (LP) problem $(P_2)$ as follows:\vspace{-0.2cm}
\begin{flalign}
\textrm{$(P_2)$:}\>\>\>\>\>\>\>\>\>\>\>\>&\min_{\boldsymbol{y}\in \mathds{R}^{H}} u_1(\boldsymbol{y})&\label{eq:P2Obj}\\
\textrm{s.t.} \>\>\>\>\>\>&\boldsymbol{M}^T\hat{\boldsymbol{y}}\leq\boldsymbol{1}_N,&\label{eq:P2C1}\\
&\hat{\boldsymbol{y}}^T\boldsymbol{1}_H=1/u_1(\boldsymbol{y}),&\label{eq:P2C2}\\
&\boldsymbol{y}=\hat{\boldsymbol{y}}u_1(\boldsymbol{y}),\,\hat{\boldsymbol{y}}\ge 0.&\label{eq:P2C3}
\end{flalign} 

As shown in~\cite[Chapter 2]{GT01}, the LP problem in~(\ref{eq:P2Obj})-(\ref{eq:P2C3}) can be reduced to the following standard maximization problem $(P_3)$: \vspace{-0.2cm}
\begin{flalign}
\textrm{$(P_3)$:}\>\>\>\>&\max_{\hat{\boldsymbol{y}}} \hat{\boldsymbol{y}}^T\boldsymbol{1}_H&\label{eq:P3Obj}\\
\textrm{s.t.} \>\>\>\>\>\>& \boldsymbol{M}^T\hat{\boldsymbol{y}}\leq\boldsymbol{1}_N,\,  
\hat{\boldsymbol{y}}\geq 0.&\label{eq:P3C1}
\end{flalign}

The solution of $(P_3)$ returns the optimal $\hat{\boldsymbol{y}}$ which can be used to calculate $u_1(\boldsymbol{y})$ as per~(\ref{eq:P2C2}). Hence, given $u_1(\boldsymbol{y})$ and $\hat{\boldsymbol{y}}$, we can compute the optimal $\boldsymbol{y}$ as per~(\ref{eq:P2C3}).

Similarly, the max-min attacker's problem can be transformed into a standard minimization problem as follows. The attacker's objective function is given by: 
\begin{align}\label{eq:AProblem}
\max_{\boldsymbol{x}\in\mathcal{X}}\min_{\boldsymbol{y}\in\mathcal{Y}}\boldsymbol{y}^T\boldsymbol{M}\boldsymbol{x}.
\end{align}

As can be seen from~(\ref{eq:AProblem}), the minimization operation is performed as a function of a given $\boldsymbol{x}$. Thus, we let 
\begin{align}\label{eq:MinU2}
u_2(\boldsymbol{x})=\min_{\boldsymbol{y}\in\mathcal{Y}} \boldsymbol{y}^T\boldsymbol{M}\boldsymbol{x}\textrm{ and } \hat{\boldsymbol{x}}=\boldsymbol{x}/u_2(\boldsymbol{x}).
\end{align}

Following similar derivations as the ones we carried out for the min-max problem (from problem $(P_1)$ to $(P_2)$ and then to $(P_3)$), the max-min problem in~(\ref{eq:AProblem}) can be reduced to the following standard minimization problem $(P_4)$:\vspace{-0.2cm}
\begin{flalign}
\textrm{$(P_4)$:}\>\>\>\>&\min_{\hat{\boldsymbol{x}}} \hat{\boldsymbol{x}}^T\boldsymbol{1}_N,&\label{eq:P4Obj}\\
\textrm{s.t.} \>\>\>\>\>\>& \boldsymbol{M}\hat{\boldsymbol{x}}\geq\boldsymbol{1}_H,\,\hat{\boldsymbol{x}}\geq 0.&\label{eq:P4C2}
\end{flalign}

The solution of $(P_4)$ returns the optimal $\hat{\boldsymbol{x}}$ which can be used to calculate $u_2(\boldsymbol{x})$ (similarly to~(\ref{eq:P2C2})): 
$\hat{\boldsymbol{x}}^T\boldsymbol{1}_N=1/u_2(\boldsymbol{x}).$ 
As a result, given the optimal $\hat{\boldsymbol{x}}$ and $u_2(\boldsymbol{x})$, we can compute the optimal $\boldsymbol{x}$ as per~(\ref{eq:MinU2}). 
 
The solutions of the LP problems $(P_3)$ and $(P_4)$ induce a \emph{mixed-strategy Nash equilibrium (NE)}, of the network interdiction game, defined next.
\begin{definition}\label{def:NE}
The strategy profile $(\boldsymbol{y}^*,\boldsymbol{x}^*)$, is an NE (equivalently a saddle-point equilibrium (SPE)) if and only if:
\begin{align}
(\boldsymbol{y^*})^T\boldsymbol{M}\boldsymbol{x^*}\leq (\boldsymbol{y})^T\boldsymbol{M}\boldsymbol{x^*}\>\> \forall \boldsymbol{y}\in\mathcal{Y},\\
(\boldsymbol{y^*})^T\boldsymbol{M}\boldsymbol{x^*}\geq (\boldsymbol{y^*})^T\boldsymbol{M}\boldsymbol{x}\> \>\forall \boldsymbol{x}\in\mathcal{X}.
\end{align}
\end{definition}

Based on the solutions of $(P_3)$ and $(P_4)$, the NE expected delivery time $T^*$ can be determined as shown in Proposition~\ref{prop:Prudent}.
\begin{prop}\label{prop:Prudent}
The solution strategies $(\boldsymbol{y}^*,\boldsymbol{x}^*)$ constitute an NE of the network interdiction game, and the solutions of LP problems $(P_3)$ and $(P_4)$ result in value functions $\mu_1(\hat{\boldsymbol{y}}^*)=(\hat{\boldsymbol{y}}^{*})^T\boldsymbol{1}_H$ and $\mu_2(\hat{\boldsymbol{x}}^*)=(\hat{\boldsymbol{x}}^*)^T\boldsymbol{1}_N$ satisfying 
$\mu_1(\hat{\boldsymbol{y}}^*)=\mu_2(\hat{\boldsymbol{x}}^*)=1/T.$   
\end{prop}
\begin{IEEEproof}
The proposed network interdiction game is a finite zero-sum game, defined over matrix $\boldsymbol{M}$, in which $U$'s and $A$'s expected payoffs, for a mixed strategy pair $(\boldsymbol{y},\boldsymbol{x})$, are given by $\Pi_A(\boldsymbol{y},\boldsymbol{x})=-\Pi_U(\boldsymbol{y},\boldsymbol{x})=\boldsymbol{y}^T\boldsymbol{M}\boldsymbol{x}=T$.
In any finite zero-sum game, if $\boldsymbol{y}'$ is a mixed security strategy for player 1 and $\boldsymbol{x}'$ is a mixed security strategy for player 2, then $(\boldsymbol{y}',\boldsymbol{x}')$ is an NE of this game~\cite{GT01}.
Thus, since $\boldsymbol{y}^*$ and $\boldsymbol{x}^*$ are mixed security strategies for the finite zero-sum network interdiction game, $(\boldsymbol{y}^*$, $\boldsymbol{x}^*)$ constitute an NE of that game. 

Given the equivalence between $(P_2)$ and $(P_3)$, and following from~(\ref{eq:P2C2}) we can derive the following: 
\begin{align}\label{eq:Prop1Eq1}
u_1(\boldsymbol{y}^*)=[(\hat{\boldsymbol{y}}^*)^T\boldsymbol{1}_H]^{-1} \Rightarrow u_1(\boldsymbol{y}^*)=1/\mu_1(\hat{\boldsymbol{y}}^*).
\end{align}

However, by definition of $u_1(\boldsymbol{y})$ and $T^*$,
\begin{align}\label{eq:Prop1Eq2}
u_1(\boldsymbol{y}^*)=\min_{\boldsymbol{y}\in\mathcal{Y}}u_1(\boldsymbol{y})=
\min_{\boldsymbol{y}\in\mathcal{Y}}\max_{\boldsymbol{x}\in\mathcal{X}}\boldsymbol{y}^T\boldsymbol{M}\boldsymbol{x}=T^*.
\end{align}

Thus, based on~(\ref{eq:Prop1Eq1}) and~(\ref{eq:Prop1Eq2}),
\begin{align}
\mu_1(\hat{\boldsymbol{y}}^*)=(\hat{\boldsymbol{y}}^*)^T\boldsymbol{1}_H=1/u_1(\boldsymbol{y}^*)=1/T. \nonumber
\end{align}

Using a similar derivation, it can be proven that   
\begin{align}
\mu_2(\hat{\boldsymbol{x}}^*)=(\hat{\boldsymbol{x}}^*)^T\boldsymbol{1}_N=1/u_2(\boldsymbol{x}^*)=1/T. \nonumber
\end{align}
\end{IEEEproof}

The proposed game might admit multiple NEs (multiple security strategies for each player). However, due to the zero-sum nature of the game, all these NEs will result in the same achieved expected delivery time~\cite{GT01}. Moreover, these NEs are interchangeable~\cite{GT01}. In other words, if $(\boldsymbol{y}^*,\boldsymbol{x}^*)$ and $(\boldsymbol{y}',\boldsymbol{x}')$ are two NEs, then, $(\boldsymbol{y}^*,\boldsymbol{x}')$ and $(\boldsymbol{y}',\boldsymbol{x}^*)$ are also NEs.  

\section{Prospect-Theoretic Analysis}\label{sec:Prospect}
In classical game theory (CGT), each player's expected payoff is calculated following expected utility theory (EUT). Based on EUT, CGT considers that a player, vendor or attacker, assesses the likelihood of achieving a certain delivery time objectively and values the merit of a pair of strategies $(\boldsymbol{y},\boldsymbol{x})$ rationally based on the \emph{expected value} of the payoff they achieve under these probabilistic strategies, as shown in~(\ref{eq:ExpectedDeliveryTime}) and equivalently in~(\ref{eq:UAVProblem}).  

However, as shown in various empirical studies and behavioral experiments~\cite{Kahneman1979Prospect,Tversky1992Cumulative}, when faced with risk and uncertainty (as is the case of our network interdiction game), the decision making processes  of individuals can significantly deviate from this full rationality, which is assumed by EUT and CGT. As such, when making decisions, humans have been found to assess outcomes, and probabilities, subjectively~\cite{Kahneman1979Prospect,Tversky1992Cumulative}. This is of particular interest to our network interdiction game for various reasons. First, the vendor or attacker can have inaccurate and disparate perceptions of the probability of success of an attack carried out at a given node. Thus, the risk level of a chosen path, or the merit of a chosen attack location can be perceived subjectively. Second, the value of the expected delivery time can be assessed, by the vendor and attacker, subjectively and differently from EUT. 
Since the merit of drone delivery systems lies in their ability of achieving very fast delivery, it is of utmost importance for the vendor to meet the delivery time, $T^{o}$, that it has promised to achieve.  For example, Amazon Prime Air promises a delivery time of \emph{less than 30 minutes}~\cite{Amazon}. Consequently, in practice, a delivery time is not assessed as an absolute quantity but relative to the reference point $T^{o}$. An increase in the expected delivery time above $T^{o}$ can be significantly detrimental to the vendor. 
For example, a delayed delivery, above 30 minutes, would cause significant Amazon Prime Air customer dissatisfaction which might lead to the failure of the UAV-delivery program. Moreover, for critical applications, such as emergency medicine delivery~\cite{DroneDeliveryLifeguardRing,DroneDeliveryDrugShipment}, very short delays can have tragic consequences. 
In this regard, one of the drawbacks of using EUT is that it perceives the calculated expected delivery time as an absolute quantity on which the vendor and attacker objectively base their chosen strategies rather than as a relative quantity, with respect to $T^{o}$, which can be valued subjectively based on the player.   
     
To this end, to accurately capture the vendor's and attacker's potential subjective perceptions, we incorporate the principles of \emph{prospect theory}~\cite{Kahneman1979Prospect,Tversky1992Cumulative} in our game. PT is a Nobel prize-winning theory which has been shown to more accurately model and predict decision makers' subjective behavior, preference, and valuations, compared to EUT. Using PT, the subjective perception of the likelihood of a probabilistic delivery time and its subjective assessment with respect to a reference point, such as $T^{o}$, can be accurately captured and modeled as shown next. 

Instead of merely calculating the expected delivery time, $T$, we focus on the valuation $V_z(T)$ for $z\in\{U, A\}$ that the vendor, $U$, or the attacker, $A$, associates with a certain $T$. Based on ~(\ref{eq:ExpectedDeliveryTime}), this valuation can be expressed as follows (for $z\in\{U,A\}$):
\begin{align}\label{eq:ValuationT}
V_z(T)\textrm{$=$}\sum_{h\in\mathcal{H}}\sum_{n\in\mathcal{N}}y_hx_n \left[v_z\Big(l_{hn}\omega_z(p_n) f^h(n)\textrm{$+$}f^h(D)\textrm{$-$}R_z\Big)\right]. 
\end{align}

In~(\ref{eq:ValuationT}), $\omega_z(.)\!:[0,1]\rightarrow\mathds{R}$ is a nonlinear weighting function and $v_z(.)\!:$ $\mathds{R}\rightarrow\mathds{R}$ is a nonlinear value function. 
The weighting function in~(\ref{eq:ValuationT}) captures the subjective perception that the vendor or attacker has of the likelihood of occurrence of probabilistic outcomes. In our network interdiction game, the outcome, when the vendor chooses path $h\in\mathcal{H}$ and the attacker chooses attack node $n\in h$, corresponds to the achieved delivery time and is probabilistic due to the underlying probabilistic success of the attack. In fact, when $U$ chooses $h$ and $A$ chooses $n\in h$, the achieved delivery time can be $(f^h(n)+f^h(D))$ with probability $p_n$ and $f^h(D)$ with probability $(1-p_n)$. 
In this regard, rather than objectively observing the probability with which each of these two outcomes can occur, each player views a weighted or distorted version of it. In this respect, player $z\in\{U,A\}$, perceives the probability that the delivery time would be equal to $f^h(n)+f^h(D)$ when $U$ chooses $h$ and $A$ chooses $n\in h$ to be equal to $w_z(p_n)$, which is a nonlinear transformation mapping of the objective probability $p_n$ to a subjective weight $w_z(p_n)$. This is known as the \emph{weighting effect}~\cite{Kahneman1979Prospect}. This transformation is defined based on various empirical studies conducted in~\cite{Kahneman1979Prospect,Tversky1992Cumulative} which have proven that, in real-life decision making, players tend to underweight high probability outcomes and overweight low probability outcomes. To accurately model the subjective probability perceptions of each player $z\in\{U,A\}$, we use the Prelec function~\cite{prelec1998probability} defined as follows (for a probability $p_n$):
\begin{align} \label{eq:weight}
w_z(p_n)=e^{-(-\ln p_n)^{\gamma_z}},\ 0<\gamma_i \le 1,
\end{align}\vspace{-0.5cm}

In addition to the weighting function, the value function in~(\ref{eq:ValuationT}) captures how the vendor and attacker value outcomes as gains and losses with respect to their reference point $R_z$ (which can, for example, correspond to $T^{o}$) rather than as absolute quantities. This is known as the \emph{framing effect}~\cite{Kahneman1979Prospect,Tversky1992Cumulative} based on which the value function of the vendor will take the following form: \vspace{-0.2cm}   
\begin{numcases}
{v_U(a_U)=}
\lambda_U(a_U)^{\beta_U}, \>\textrm{if}\> a_U\geq 0, \nonumber \\\label{eq:ValueFc}
-(-a_U)^{\alpha_U}, \>\textrm{if}\> a_U< 0,
\end{numcases}
\vspace{-0.3cm}
\begin{align}
\textrm{where }\>\>a_U=l_{hn}\omega_U(p_n) f^h(n)+f^h(D)-R_U,
\end{align}
while $\lambda_U$, $\beta_U$, and $\alpha_U$ are positive constants (with $\lambda_U>1$) and $\omega_U(.)$ is as given in~(\ref{eq:weight}).
In fact, since the vendor is a minimizer, $a_U\geq0$ correspond to losses and $a_U<0$ corresponds to gains. This value function captures the following PT-specific properties: i) the value that the vendor associates with a certain delivery time is assessed as a gain or loss with respect to a subjective reference point $R_U$ (e.g. $T^o$) rather than as an absolute quantity, and ii) losses loom larger than gains, as measured by the loss multiplier $\lambda_U$ in~(\ref{eq:ValueFc}), which captures the fact that the vendor amplifies the effect of crossing the promised delivery time. 
For the attacker, a similar expression for the value function as in~(\ref{eq:ValueFc}) can be used while adjusting for the fact that the attacker is a maximizer:\vspace{-0.2cm}
\begin{numcases}
{v_A(a_A)=}
-\lambda_A(-a_A)^{\beta_A}, \>\textrm{if}\> a_A< 0, \nonumber \\
(a_A)^{\alpha_A}, \>\textrm{if}\> a_A\geq 0,
\end{numcases}
\vspace{-0.3cm}
\begin{align}
\textrm{where}\>\>a_A=l_{hn}\omega_A(p_n) f^h(n)+f^h(D)-R_A.
\end{align}

In addition, to incorporate PT in our network interdiction game, we define the $(H\times N)$ matrices $\boldsymbol{M}^{U,\textrm{PT}}$ and $\boldsymbol{M}^{A,\textrm{PT}}$ whose elements are, respectively, given by $(\forall h\in\mathcal{H}\,, \forall n\in\mathcal{N})$: 
\begin{align}
m^{U,\textrm{PT}}=v_U\left(l_{hn}\omega_U(p_n) f^h(n)\textrm{$+$}f^h(D)\textrm{$-$}R_U\right),\\
m^{A,\textrm{PT}}=v_A\left(l_{hn}\omega_A(p_n) f^h(n)\textrm{$+$}f^h(D)\textrm{$-$}R_A\right).
\end{align}\vspace{-0.6cm}                          

As such, to choose its mixed path-selection strategy, the vendor must solve the following optimization problem, $(P_5)$:\vspace{-0.2cm} 
\begin{align}\label{eq:P5Obj}
\min_{\boldsymbol{y\in\mathcal{Y}}}\max_{\boldsymbol{x\in\mathcal{X}}} \boldsymbol{y}^T\boldsymbol{M}^{U,\textrm{PT}}\boldsymbol{x}. 
\end{align}\vspace{-0.4cm}

On the other hand, the defender solves the following optimization problem, $(P_6)$:\vspace{-0.2cm}
\begin{flalign}\label{eq:P6Obj}
\max_{\boldsymbol{x\in\mathcal{X}}}\min_{\boldsymbol{y\in\mathcal{Y}}} \boldsymbol{y}^T\boldsymbol{M}^{A,\textrm{PT}}\boldsymbol{x}.
\end{flalign}\vspace{-0.4cm}

In practice, neither the vendor nor the attacker will have full knowledge about the subjectivity level of their opponent. Hence, a common practice in security settings~\cite{SanjabSaadJournal} is for each player to consider that the opponent will always choose the strategy that inflicts the worst consequence on this player. This property has been captured, respectively, by the min-max and max-min formulations of $(P_5)$ and $(P_6)$. 
Problem $(P_5)$ and $(P_6)$ can be reduced, respectively, into standard maximization and minimization problems following a similar transformation process as the one described in Section~\ref{sec:Game}. As opposed to the analysis in Section~\ref{sec:Game}, however, here we will not have an SPE since $\boldsymbol{M}^{U,\textrm{PT}}$ and $\boldsymbol{M}^{A,\textrm{PT}}$ are different and, in such cases, security strategies for the players do not generally lead to an SPE~\cite{GT01}.  
 
\section{Numerical Results}\label{sec:NumRes}

For our numerical simulations, we consider a directed graph with $N=10$ nodes and $E=18$ edges as shown in Fig.~\ref{fig:Network}. 
We choose $[t_1, t_2,...,t_{18}]$ $\triangleq$ $[3,$ $3,$ $3,$ $6,$ $6,$ $3,$ $6,$ $6,$ $6,$ $8,$ $6,$ $8,$ $10,$ $10,$ $10,$ $14,$ $12,$ $14]$ and $[p_1,$ $p_2,$$...,$$p_{10}]$ $\triangleq$ $[0,$ $0.2,$ $0.4,$ $0.2,$ $0.4,$ $0.4,$ $0.5,$ $0.8,$ $0.5,$ $0]$. 
We number the paths as follows: $[1,$ $2,$ $...,$ $18]$ $\triangleq$ $[(2,5,7),$ $(2,5,8),$ $(2,5,9),$ $(2,6,7),$ $(2,6,8),$ $(2,6,9),$ $(3,5,7),$ $(3,5,8),$ $(3,5,9),$ $(3,6,7),$ $(3,6,8),$ $(3,6,9),$ $(4,5,7),$ $(4,5,8),$ $(4,5,9),$ $(4,6,7),$ $(4,6,8),$ $(4,6,9)]$ where, since node 1 ($O$) and node 10 ($D$) are part of all paths, a path $(i,j,k)$ corresponds to $(1,i,j,k,10)$ . Moreover, for the PT parameters, unless stated otherwise, we take $\lambda_A=\lambda_U=5$, $\beta_U=\beta_A=0.8$, and $\alpha_U=\alpha_A=0.2$.
 
\begin{figure}[t!]
  \begin{center}
   \vspace{-0.35cm}
    \includegraphics[width=7cm]{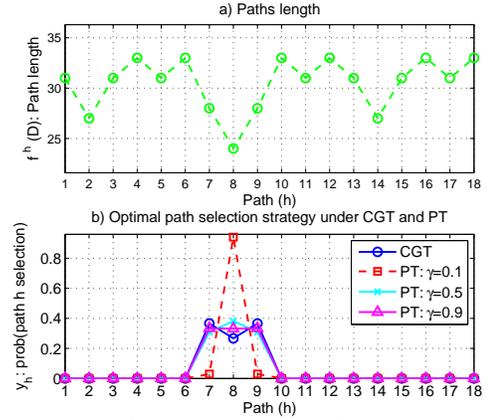}
    \vspace{-0.4cm}
    \caption{\label{fig:VendorAlpha} a) Path length for each path in $\mathcal{H}$, b) Optimal path selection strategy under CGT and PT for various values of the rationality parameter.}
  \end{center}\vspace{-0.3cm}
\end{figure}
\begin{figure}[t!]
  \begin{center}
   \vspace{-0.35cm}
    \includegraphics[width=7cm]{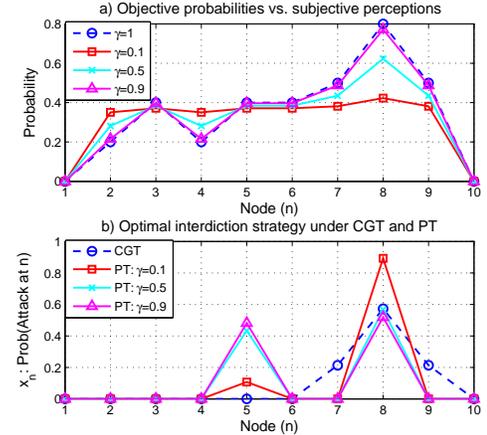}
    \vspace{-0.4cm}
    \caption{\label{fig:AttwrtAlpha} a) Objective and subjective perceptions of $p_n$, b) Optimal interdiction strategy under CGT and PT -- for various values of the rationality parameter.}
  \end{center}\vspace{-0.9cm}
\end{figure}
First, in Fig.~\ref{fig:VendorAlpha}a, we show the length from $O$ to $D$ for each of the possibles paths in $\mathcal{H}$. This figure shows that the shortest path is path $8$ followed by paths $2$ and $14$.  

Fig.~\ref{fig:VendorAlpha}b shows the optimal path strategy chosen by the vendor. Clearly, under CGT, the shortest path (path 8) is not the one that is chosen with the highest probability. In fact, the vendor is more likely to choose either path $7$ or $9$ due to the fact that path $8$ is risky since $p_8=0.8$. 
However, under PT, the weighting effect flattens the perceived probabilities as shown in Fig.~\ref{fig:AttwrtAlpha}a. In this regard, Fig.~\ref{fig:AttwrtAlpha}a shows the objective probability $p_n$ at each $n\in\mathcal{N}$ and the distorted weighted versions of these probabilities for different values of the rationality parameter $\gamma$, where in this case $\gamma=\gamma_U=\gamma_A$. This figure highlights the under-weighting of high probabilities ($p_n>0.4$) and the over-weighting of low probabilities, based on which, a very irrational vendor, $\gamma=0.1$, perceives the probability of a successful attack to be almost equally likely at all nodes between $O$ and $D$. 
Consequently, as shown in Fig.~\ref{fig:VendorAlpha}b, under PT, the defender becomes more likely to take the shortest path. At this extreme level of rationality, for $\gamma=0.1$, since the vendor perceives the probability of a successful attack to be equal among all nodes, the vendor assesses all paths to be equally risky and hence chooses the shortest path with probability 0.94.

Fig.~\ref{fig:AttwrtAlpha}b shows the optimal interdiction strategy of the attacker under CGT and PT, for different values of $\gamma$. Under CGT, the attacker will optimally choose to randomize between nodes 7, 8, and 9 with the highest probability of launching the attack at node 8, knowing that node 8 is part of the shortest path and that $p_8=0.8$. However, under PT, the attacker focuses its attack on nodes $5$ and $8$ which are part of the shortest paths. 

\begin{figure}[t!]
  \begin{center}
   \vspace{-0.35cm}
    \includegraphics[width=7cm]{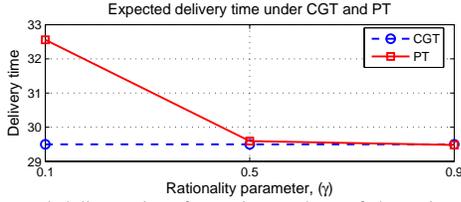}
    \vspace{-0.4cm}
    \caption{\label{fig:ExpectedDeliverywrtGamma} Expected delivery time for various values of the rationality parameter, $\gamma$.}
  \end{center}\vspace{-0.3cm}
\end{figure}

From Figs.~\ref{fig:VendorAlpha} and~\ref{fig:AttwrtAlpha}, we can see that the weighting effect, and particularly, the rationality parameter, have a very impactful effect on the chosen path and attack strategies, hence, significantly affecting the expected delivery time. In fact, Fig.~\ref{fig:ExpectedDeliverywrtGamma} shows the variation in the achieved expected delivery time for $\gamma\in\{0.1, 0.5, 0.9\}$. For instance, lower rationality levels lead to higher delivery times. In fact, when $\gamma$ decreases from 0.9 to 0.1, the achieved expected delivery time increases by 11\%. Moreover, in this figure, we consider the target delivery time, $T^{o}$ to be such that $T^{o}=R_U=R_A=30$. 
Thus, the distorted perception of probability leads to choosing risky path strategies which achieve expected delivery times that exceed the target time. This can have dire concequences especially in critical and emergency medicine delivery applications~\cite{DroneDeliveryLifeguardRing,DroneDeliveryDrugShipment}. Here, we note that our calculated delivery time actually corresponds to the expected \emph{flight time} of the UAV when faced with attacks. The actual delivery time will include additional processing times which can be mathematically modeled as additive constants. Thus, a successful attack incurs additional delays since re-sending a replacement product requires additional time for re-processing and re-handling.        

\begin{figure}[t!]
  \begin{center}
   \vspace{-0.35cm}
    \includegraphics[width=7cm]{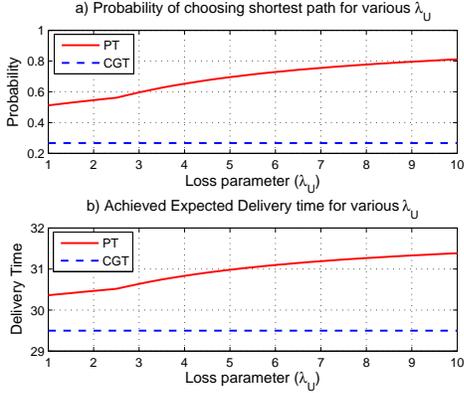}
    \vspace{-0.4cm}
    \caption{\label{fig:ShortestandTimewrtlambda} Variation with respect to $\lambda_U$ of a) Probability of choosing the shortest path, b) Achieved expected delivery time.}
  \end{center}\vspace{-0.9cm}
\end{figure}

Fig.~\ref{fig:ShortestandTimewrtlambda} investigates the effect of the loss parameter $\lambda_U$ on the probability of choosing the shortest path and on the achieved expected delivery time, for $R_U=30$.   
In fact, an increase in $\lambda_U$ reflects that a player exaggerates losses further and hence is more averse to losses. 
For our game, when $\lambda_U$ increases, the vendor significantly exaggerates the consequences of not meeting the target delivery time and, hence, is more likely to choose risky path strategies that have shorter path lengths. Indeed, as shown in Fig.~\ref{fig:ShortestandTimewrtlambda}a, the vendor is significantly more likely to choose the shortest path when $\lambda_U$ increases. In fact, when $\lambda_U$ increases from 1 to 10, the probability of choosing the shortest path increases from 0.51 to 0.81. This risky path selection strategy will have a negative effect in terms of the expected delivery time. Indeed, Fig.~\ref{fig:ShortestandTimewrtlambda}b shows that the expected delivery time increases with $\lambda_U$. An important observation here is that, under the subjective behavior observed by PT, the expected delivery time exceeds that under CGT as well as the target delivery time. Hence, this shows that the subjective perception of probabilities and outcomes by the vendor can impair its chosen path strategies incurring delays to the delivery time. \vspace{-0.2cm}
 
\section{Conclusion}\label{sec:Conclusion}
In this paper, we have introduced a novel mathematical framework for studying the cyber-physical security of drone delivery systems against interdiction attacks. We have modeled the problem using a zero-sum network interdiction game between a vendor and an attacker. In this regard, we have proven that the Nash equilibrium of the game can be obtained using the solution of standard linear programming problems. In addition, to capture the subjective behavior of the vendor and the attacker, we have incorporated the notions of prospect theory in our game formulation. Simulation results have shown that the subjective decision making processes of the vendor and attacker lead to delays in delivery time which can surpass the target delivery time to which the vendor has committed.

\def\baselinestretch{0.82}
\bibliographystyle{IEEEtran}
\bibliography{reference}

\end{document}